\documentclass[useAMS,usenatbib]{mnras}

\usepackage{graphicx}
\usepackage{epstopdf}
\usepackage{amsmath,esint,bm}

\usepackage{etoolbox}
\makeatletter
\patchcmd\@combinedblfloats{\box\@outputbox}{\unvbox\@outputbox}{}{%
   \errmessage{\noexpand\@combinedblfloats could not be patched}%
}%
 \makeatother

\title[A Varying $G$ Between MOND and Weak Weyl Gravity]{Interposing a Varying Gravitational Constant Between Modified Newtonian Dynamics and Weak Weyl Gravity}

\author[Christodoulou \& Kazanas]{Dimitris M. Christodoulou$^{1,2}$  and Demosthenes Kazanas$^{3}$
\\
$^{1}$Lowell Center for Space Science and Technology, University of Massachusetts Lowell, Lowell, MA, 01854, USA.\\
$^{2}$Dept. of Mathematical Sciences, Univ. of Massachusetts Lowell,
Lowell, MA, 01854, USA. E-mail: dimitris\_christodoulou@uml.edu\\
$^{3}$NASA/GSFC, Laboratory for High-Energy Astrophysics, Code 663, Greenbelt, MD 20771, USA. E-mail: demos.kazanas@nasa.gov\\
}

\begin{document}

\def\gsim{\mathrel{\raise.5ex\hbox{$>$}\mkern-14mu
                \lower0.6ex\hbox{$\sim$}}}

\def\lsim{\mathrel{\raise.3ex\hbox{$<$}\mkern-14mu
               \lower0.6ex\hbox{$\sim$}}}

\pagerange{\pageref{firstpage}--\pageref{lastpage}} \pubyear{2018}

\maketitle

\label{firstpage}

\begin{abstract}
The Newtonian gravitational constant $G$ obeys the dimensional relation $[G] [M] [a] = [v]^4$, where $M$, $a$, and $v$ denote mass, acceleration, and speed, respectively. Since the baryonic Tully-Fisher (BTF) and Faber-Jackson (BFJ) relations are observed facts, this relation implies that $G\, a = {\rm constant}$. This result cannot be obtained in Newtonian dynamics which cannot explain the origin of the BTF and BFJ relations. An alternative, modified Newtonian dynamics (MOND) assumes that $G=G_0$ is constant in space and derives naturally a characteristic constant acceleration $a=a_0$, as well as the BTF and BFJ relations. This is overkill and it comes with a penalty: MOND cannot explain the origin of $a_0$. A solid physical resolution of this issue is that $G \propto a^{-1}$, which implies that in
lower-acceleration environments the gravitational force is boosted relative to its Newtonian value because $G$ increases. This eliminates all problems related to MOND's empirical cutoff $a_0$ and yields a quantitative method for mapping the detailed variations of $G(a)$ across each individual galaxy as well as on larger and smaller scales. On the opposite end, the large accelerations produced by $G(a)$ appear to be linked to the weak-field limit of the fourth-order theory of conformal Weyl gravity.
\end{abstract}




\section{Introduction and Motivation}\label{intro}

\subsection{Weyl Gravity and Objections to Dark Matter}

Dark matter (DM) has been a staple of astrophysics since \cite{zwi33,zwi37} realized that the observed luminous matter of the Coma cluster is not sufficient to support the observed velocities of member galaxies. Its historical record is enormous and diverse, which is ironic for an untangible medium that has never been detected directly, even by the many ongoing experiments that continue to report routinely one failure to detect after another \citep{rott17}. This 40-year stubborn belief that this new aether actually does exist reminds us of the aether of \cite{huy90} that held back progress in physics for more than 200 years. It also rests with the belief that the Newtonian potential is truly the weak-field limit of the ``correct" theory of strong gravity, namely General Relativity (GR).

The most concrete case for DM has been made so far by the ``flat" (i.e., independent of the galactocentric distance) rotation curves of spiral galaxies at distances with very little luminous matter, where they should instead be decreasing in accordance with Kepler's third law. Because of their well-defined dynamics and the reduced observational ambiguity of their kinematics, these systems have also allowed for detailed determinations of the spatial distribution of DM in all major types of galaxies.

Since the beginning, the assumption that DM exists has not been taken well by a number of researchers for various reasons. Our objection to this aetherial idea is simple: in trying to fit galaxy rotation curves with DM, several new free parameters are introduced (shape, size, density profile, several different radii for maximum rotation, shape truncation, etc.), not just one as is naively believed. Even more parameters need to be introduced in modeling various features of the rotation curves such as downturns in the inner regions and upturns at the outermost regions. 

It is not surprising then that some researchers have devoted their last 40 years searching for alternatives that rely on only one free parameter, as solid physical reasoning would dictate and expect. The best known alternative to DM is Modified Newtonian Dynamics (MOND) \citep{mil83a,mil83b,mil83c,fam12,mil15c}. The crucial feature of this proposal is the introduction of {\it one and only one} free parameter, the 
characteristic acceleration $a_0$ which modifies Newton's second law for accelerations $a \ll a_0$; so, at regions where  $a > a_0$, the dynamics is given by Newton's second law $F = m a$ ($F$ is force and $m$ is mass); whereas for $a \ll a_0$, the equation becomes $F = m \sqrt{a a_0}$ with $a_0 \simeq 10^{-8}$ cm s$^{-2}$. Noting that $a_0 \simeq c H_0 \simeq c/T_U$, with $c$ the speed of light, $H_0$ the Hubble constant, and $T_U$ the age of the universe, these conspiring relations suggest that the dynamics of galaxies (and all systems in which $a \ll a_0$) may be influenced by physics at the cosmological scale, an issue clearly at odds with GR. But this could be just a numerical coincidence.

Another alternative to DM is Weyl Gravity (WG). This theory has been investigated by \cite{man89} as a potential alternative to GR. The interest in WG stems from the desirable properties it possesses and that are absent from GR. In addition to general covariance, a local scale invariance property is also present in the theory (a symmetry shared by all other fundamental interactions and by MOND). These conditions suffice to make the action of the theory be the square of the Weyl tensor $C^{\kappa\lambda\mu\nu}$, viz.
$$
{\cal S}_{WG} = \alpha\int C^{\kappa\lambda\mu\nu}C_{\kappa\lambda\mu\nu}d^4 x\, , 
$$
where $\alpha$ is the dimensionless coupling constant of WG. With respect to this action, one should note that the resulting equations for the gravitational field are now of fourth order, and far more importantly, Newton's gravitational constant $G$ is absent. Despite their higher order, the WG equations of the vacuum, static, spherically symmetric geometry have an exact solution given by
\citep[][]{man89}
$$
g_{00}=\frac{1}{g_{rr}} = 1 - 3\beta\gamma -\frac{\beta (2 - 3 \beta\gamma)}{r} + \gamma r - k r^2\, ,
$$
where $\beta, \gamma, k$ are integration constants. While the $1/r$ and the $r^2$ terms are recognized as the Schwarzschild and cosmological curvature terms of GR, the linear term $\gamma r$, representing a constant acceleration of order $a_0=c H_0$, is a brand new term not encountered in the well-known vacuum solutions of GR. Being of the same order as that of the acceleration introduced by MOND, it is not surprising that inclusion of this term in the geodesic equations provides reasonable fits to the galactic rotation curves \citep[][]{man11,man12}.

Therefore, what makes WG a compelling theory is that its vacuum solution engenders an acceleration of the order required by the dynamics and observed kinematics of galaxies, although no such demand was built into the action integral: its origin is simply the result of local scale invariance of the theory \citep[the same symmetry that appears also in the deep MOND limit;][]{mil15c}. Despite the theoretical and observational impetus, the absence of $G$ from the WG action integral makes it difficult to apply WG to customary observations. It is reasonable to imagine that the fourth-order equations will somehow allow/provide for an effective $G$ that could be involved in modeling of real astrophysical systems. 
But such an effective $G$ does not have to be a constant. It could exhibit a variation in space, time, or both.

Below we show that this is indeed the case for the $G$ of Newtonian dynamics. The familiar $G$ varies in space. It is mandated to be a function of acceleration $a$ and its variation produces correctly the deep MOND limit and the weak-field WG limit.

\subsection{An Investigation of G}

The Newtonian gravitational constant $G$ has dimensions of
\begin{equation}
[G] = \frac{[L]^3}{[M]\, [T]^2} = \frac{[L]/[T]^2}{[M]/[L]^2} = \frac{[a]}{[\Sigma]} \, ,
\label{dim1}
\end{equation}
where $a$ is acceleration and $\Sigma$ is surface density. We refrain from interpreting
this dimensional relation because $a$ and $\Sigma$ are not fundamental variables, they
are instead derived from fundamental variables. Nevertheless, there have been many
reports for and against a constant $\Sigma=\Sigma_0$ across all astrophysical scales in
visible and dark matter
\citep[][]{lar81,kaz95,hey09,don09,gen09,lom10,bal12,del12,del13, mil16,tra18}; and
even more reports of a constant $a=a_0$ in the rotation curves of spiral galaxies, in
dwarf and elliptical galaxies, in clusters of galaxies
\citep[][]{mil83a,mil83b,mil83c,man89,san02,fam12,mcg13,mil15a,mil15b,mil17}, and,
quite surprisingly, in the Oort cloud \citep{pau17}.

Using dimensional analysis, we rewrite eq.~(\ref{dim1}) in the equivalent form
\begin{equation}
[G] [M] [a] = [v]^4 \, ,
\label{dim2}
\end{equation}
where $v$ is speed. This relation is not fundamental either (it contains two
derivatives, $a$ and $v$), but it can be interpreted correctly with help from
observations. Observations have established the baryonic Tully-Fisher (BTF) relation in
spiral galaxies \citep{tul77,mcg00,mcg12} and the baryonic Faber-Jackson (BFJ) relation
in elliptical galaxies \citep{fab76,san09,den15}; and these relations imply for the
visible matter that $M\propto v^4$. Then, eq.~(\ref{dim2}) demands that
\begin{equation}
G\, a = {\rm constant}\, ,
\label{dim3}
\end{equation}
a result that is fundamental for our understanding of gravity and that explains right
away the origin of MOND's elusive constant ${\cal A}_0\equiv G_0a_0$ \citep{mil15c}.
Eqs.~(\ref{dim1}) and~(\ref{dim3}) also imply that $a\propto\Sigma^{1/2}$ and
$G\propto\Sigma^{-1/2}$. Thus, the reported variations of $\Sigma$ on various
astrophysical scales (especially in Giant Molecular Clouds) could potentially be
tracing various differing values of $G(a)$.

Conversely, if one accepts the validity of eq.~(\ref{dim3}), then the BTF and BFJ
relations are naturally explained. There is no need to assume separately that $G$ and
$a$ exhibit universal constants, only that their product must be constant, which defines only one constant, ${\cal A}_0= G_0a_0$, in the deep MOND limit. And this is
where we deviate from MOND in this work: MOND assumes
implicitly  that $G=G_0$ is constant in space, and then eq.~(\ref{dim3}) naturally
predicts the existence of a constant acceleration $a=a_0$ \citep[see, e.g., the recent
results of][]{mcg16,lel17}.

\begin{figure}
\begin{center}
    \leavevmode
      \includegraphics[trim=0 0cm 0 0.1cm, clip, angle=0,width=9 cm]{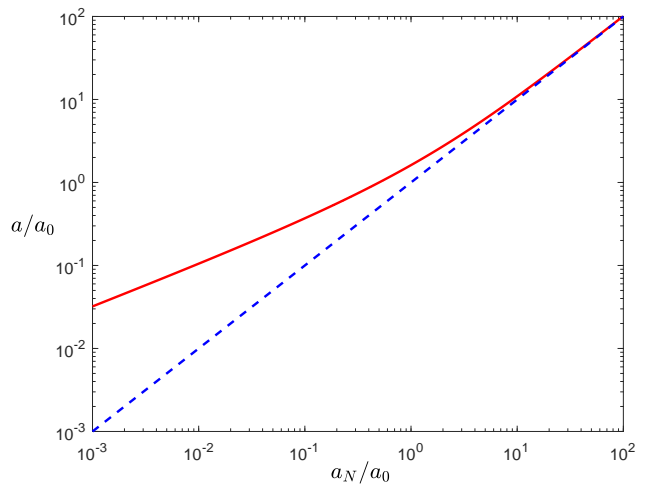}
\caption{Acceleration $a$ (solid line) is plotted against its Newtonian value $a_N$
(dashed line) using eq.~(\ref{dim5}). The same type of diagram is also borne out
from vast observations of 2693 galaxies \citep{mcg16,lel17}.
\label{fig1}}
  \end{center}
\end{figure}

In the following, we make no assumptions as to the constancy of $G$ or $a$
individually; instead we accept only what eq.~(\ref{dim3}) tells us, which is that, at
very low accelerations,
\begin{equation}
G \propto a^{-1}.
\label{dim4}
\end{equation}
Thus, $G$ varies {\it inversely  with acceleration} $a$ and the gravitational force
will get a larger boost than its Newtonian value in lower-acceleration environments,
such as the outer fringes of our solar system and the outer fringes of galaxies. This
boost is shown in Fig.~\ref{fig1}, where $a$ (solid line) stays higher than its
Newtonian value $a_N$ (dashed line) for $a_N \leq 10\,a_0$.

Cosmological speculations aside, MOND's constant acceleration $a_0$ cannot be presently
justified on physical grounds. Some of the advantages of having $a_0$ appear as a
scaling parameter in the spatial variation of $G(a)$ are the following:
\begin{itemize}
\item[1.]A continuously increasing $G(a)$  function with decreasing $a$ alleviates
    the need for finding an explanation for a solitary ``fundamental'' constant
    $a_0$.
\item[2.]The dynamics in the vicinity of $a\sim a_0$ is no longer unspecified (see
    eq.~(\ref{dim5}) below) and there is no need to introduce ``interpolating
    functions'' such as those arbitrarily assumed in MOND for the gravitational
    force.
\item[3.]As a consequence of item 2, the rotation curves in the inner few kiloparsecs
    of galaxies do not require fine tunning, especially those that decline toward
    ``asymptotic flatness'' {\it from above}.
\item[4.]The constant $a_0$ appears to have some minor effect in the Newtonian  limit of high accelerations, therefore it is present in the entire domain of accelerations (see \S~\ref{vG} and \S~\ref{wG} below).
\item[5.]There is no need to modify the inertial mass $m$ in Newton's second law
    $F=m\,a$, so the Weak Equivalence Principle remains valid.
\item[6.]The Strong Equivalence Principle (SEP) is invalid since $G$ varies in space. This provides a lifeline to many advanced cosmological theories that do not satisfy the SEP.
\item[7.]A varying $G(a)$ paves the way for future investigations of the
    gravitational field (and, incidentally, the electrostatic field) at very low
    accelerations \citep{sul17}, where it may not be falling as $1/r^2$---a problem
    that has not yet received much attention.
\end{itemize}
In summary, we believe that no challenge currently posed to MOND in its deep limit
\citep{chr88,del12,del13,mil13,mil15a,mil17,mil18,fam18} can survive if
eq.~(\ref{dim4}) is adopted as a first principle instead of the familiar but arbitrary
constant acceleration $a_0$.

\subsection{Outline}

In what follows, we discuss a varying $G(a)$ equation and its limiting behavior for
high and low accelerations (\S~\ref{vG}). Then we discuss the relationship between the
high accelerations and conformal Weyl gravity (\S~\ref{wG}). We present some concluding
remarks in \S~\ref{conc}. In Appendices, we solve Poisson equations with a varying $G(a)$, in order to derive and compare the interior potentials
in a sphere of uniform mass density.

\section{Varying Gravitational Constant}\label{vG}

The simplest elementary function $G(a)$ that exhibits correct asymptotic behavior at
both high and low accelerations is the sum of the two asymptotic values,\footnote{For $a_0\ll a$, then $G\to G_0$. In the opposite
limit, $a_0 \gg a$, we recover eq.~(\ref{dim4}) and MOND's deep limit. Any other choice
of $G(a)$ with the same two asymptotic behaviors could introduce additional spurious
terms to the equations; and if did not, it would have to be rejected on the basis of simplicity by Occam's razor.} viz.
\begin{equation}
G = G_0\left(1 + \frac{a_0}{a}\right) \, .
\label{g}
\end{equation}
Using this equation, we find for the acceleration that
\begin{equation}
a = \frac{GM}{r^2} = \frac{a_N}{2}\left( 1 + \sqrt{1 + \frac{4a_0}{a_N}}\, \right) \, ,
\label{dim5}
\end{equation}
where $a_N = G_0 M/r^2$ is the  Newtonian acceleration in the gravitational field of
mass $M$ at an exterior distance $r$ from the center, and $G_0$ and $a_0$ are the
presently accepted values of these constants. We distinguish two limiting cases:
\begin{itemize}
\item[(a)]In the Newtonian limit $a_N \gg a_0$, we obtain
\begin{equation}
a \approx a_N + a_0\left[ 1 - ({a_0}/{a_N}) + 2({a_0}/{a_N})^2 \right]  \, ,
\label{dim5a}
\end{equation}
which is the Newtonian acceleration  along with higher-order correction terms. Only
in the limit of $a_0\to 0$ do we recover the classical Newtonian acceleration $a_N$,
as we should. In this limit, $G(r=0)= G_0$ and $G(r)=G_0+(a_0/M)r^2$, i.e., $G(r)$
increases quadratically with distance $r$.
\\
\item[(b)]In the deep MOND limit $a_N \ll a_0$, we obtain
\begin{equation}
a \approx \sqrt{a_0\, a_N} + \frac{a_N}{2}\left[ 1 + ({{a_N}/{a_0}})^{1/2}/4  - ({a_N}/{a_0})^{3/2}/64 \right]\, ,
\label{dim5b}
\end{equation}
which is essentially MOND's geometric-mean  acceleration plus additional correction
terms. In this limit, $G(r)=\sqrt{G_0a_0/M}\,r + G_0/2$, i.e., $G(r)$ increases
linearly with $r$.
\end{itemize}

\section{Weak-Field Weyl Gravity}\label{wG}

Here we focus on the high-acceleration regime, eq.~(\ref{dim5a}). We restore the
Newtonian value of $a_N$ outside a mass $M$, viz.
\begin{equation}
a_N =\frac{G_0 M}{r^2} \, ,
\label{weyl1}
\end{equation}
where $r$ is the distance from the center, and we obtain
\begin{equation}
a = \frac{G_0 M}{r^2} + a_0 - \frac{a_0^2\, r^2}{G_0 M} + {\cal O}(a_0^3\,r^4) \, .
\label{weyl2}
\end{equation}
The first two terms on the right-hand side match the Weyl acceleration $a_{_{WG}}=2\beta/r^2 +
\gamma$ (where $\beta$, $\gamma = {\rm const.}$) found by \cite{man89}, but the next
term ($\sim a_0^2\,r^2$) is of higher order than the cosmological Weyl acceleration
$-2kr$, where $k$ is the curvature of spacetime. This tells us that the correction
terms in eq.~(\ref{weyl2}) do not include cosmological influences or effects; thus,
$G_0$ and $a_0$ appear to just be local power-law constants of eq.~(\ref{g}) and they
may even be found to vary from galaxy to galaxy \citep[since only the product ${\cal
A}_0=G_0a_0$ is required to be constant in the deep MOND limit; see eq.~(\ref{dim3})
above as well as Sec. IV in][]{mil15c}; and, perhaps, in time as well
\citep{bar96,kp18}.

Terms of order $a_0^2\,r^2$ and higher cannot be produced by fourth-order Weyl gravity, but they can be produced by higher-even-order conformal theories \citep{man94}. Therefore, the variation of $G(a)$ in the high-acceleration regime does not produce identically the weak-field terms of Weyl gravity (this tells us that eq.~(\ref{g}) may exhibit richer phenomenology than the weak-field limit of Weyl gravity). It could then be that the ``correct'' relativistic theory of gravity is a conformal theory of order higher than four.

In the series expansion shown in eq.~(\ref{weyl2}), notice that the $1/r$ term is also missing. This means that there is no logarithmic term in the exterior solution of the gravitational potential $\Phi(r)$. This is in agreement with the Weyl vacuum solution found by \cite{man89} and it precludes the presence of DM-like terms (with $\Phi_{\rm DM}\propto\ln r$) in systems in which $G(a)$ varies with acceleration according to eq.~(\ref{g}). So DM and a varying $G(a)\propto a^{-1}$ appear to be mutually exclusive propositions. In this context, we reiterate that, as of summer 2017, DM searches are not doing well \citep{rott17}, despite many extensive and expensive funding campaigns. Thus, the time is ripe---and right---for investigations to turn to variations of $G(a)$ in galaxies and galaxy clusters.

\section{Concluding Remarks}\label{conc}

We have shown in simple terms that the BTF and BFJ relations imply that the gravitational constant varies with acceleration as $G\propto a^{-1}$ in MOND's low-acceleration regime of galaxy and galaxy cluster dynamics. Conversely, the hypothesis that $G\propto a^{-1}$ in the deep MOND limit leads immediately to the justification of these long-known relations \citep{fab76,tul77}. MOND phenomenology can also produce the same relations by assuming that $G=G_0$ and $a=a_0$ are individual universal constants at galactic scales and beyond (but these are two ad hoc assumptions, not one). In the deep MOND limit, we only need a single constant ${\cal A}_0\equiv G_0\,a_0$ \citep[see also][]{mil15c}.

We work with a single new modification of gravity as indicated by eq.~(\ref{g}) above. What makes this equation attractive is its simplicity and its limiting cases: In the deep MOND limit of small accelerations, our total acceleration reduces to and justifies MOND accelerations; in the Newtonian limit of large accelerations, our total acceleration matches the noncosmological terms of conformal WG in its weak-field approximation; and in intermediate accelerations $a\sim a_0$, eqs.~(\ref{g}) and~(\ref{dim5}) provide much needed continuity across the entire range of accelerations.

Such precise agreement between accelerations in the two limiting cases cannot possibly be coincidental. It rather signifies that a varying $G(a)$ describes and reproduces weak-field WG and MOND,  two gravitational paradigms which strive to describe dynamics in galaxies and beyond in the absence of dark matter. Eqs.~(\ref{g}) and~(\ref{dim5}) constitute a first step toward a new theory of modified dynamics and eq.~(\ref{weyl2}) possibly gives us an idea of a relativistic conformal extension (MOND is also conformal in its deep limit) whose coupling constant in the action integral is conveniently unitless. This property has been long sought for gravity, so that the action of this last fundamental force of nature may finally resemble the other four forces whose coupling constants are all unitless \citep{man89}.

\section*{Acknowledgments}
We thank anonymous referees for suggestions that clarified considerably the presentation of these ideas. NASA support over the years is also gratefully acknowledged.

\appendix
\section{Interior Solution of Nonlinear Spherical Poisson Equation in the Deep MOND Limit}

Consider a spherical ($r$) homogeneous distribution of matter with density $\rho_0$ in the deep MOND limit where $a<<a_0$ and $G=G_0\,a_0/a$ (eq.~(\ref{dim4})). We are interested in the interior solution for the gravitational potential $\Phi(r)$. Then Poisson's equation takes the nonlinear form
\begin{equation}
\left(\frac{d\Phi}{dr}\right)\left[\frac{1}{r^2}\frac{d}{dr}\left(r^2\frac{d\Phi}{dr}\right)\right] = S \, ,
\label{p1}
\end{equation}
where $S=4\pi G_0\rho_0 a_0 = {\rm constant}$ inside the sphere.

Here we determine the interior solution $\Phi(r)$ in the form of a power law in radius, viz.
\begin{equation}
\Phi(r) = A\, r^n\, ,
\label{p2}
\end{equation}
where $A$ and $n$ are positive constants and the boundary condition is $\Phi(0)=0$. Substituting eq.~(\ref{p2}) into eq.~(\ref{p1}), we find that
\begin{equation}
A^2\,n^2(n+1)\,r^{2n-3} = S\, .
\label{p3}
\end{equation}
A self-consistent solution then has $n=3/2$ and
\begin{equation}
A = (2/3)\sqrt{(2/5)S} \, ;
\label{p4}
\end{equation}
and then eq.~(\ref{p2}) takes the form
\begin{equation}
\Phi(r) = (4/3)\sqrt{(2/5)\pi G_0\rho_0 a_0} ~ r^{3/2}\, .
\label{p5}
\end{equation}
This solution is well-behaved at the origin and assumes its maximum value on the surface of the sphere (say, $r=r_0$), where it is expected to match the exterior solution. We note that Gauss's law does not apply in this case, yet its naive application deduces the correct radial behavior of $\Phi\propto r^{3/2}$.

\section{Interior Solution of Nonlinear Spherical Poisson Equation in the Newton-Weyl Limit}

Consider again a spherical ($r$) homogeneous distribution of matter with density $\rho_0$ in the limit where $a>>a_0$ and $G=G_0(1+a_0/a)$ (eq.~(\ref{g})). We are interested in the interior solution for the gravitational potential $\Phi(r)$. Then Poisson's equation takes the nonlinear form
\begin{equation}
\left(\frac{d\Phi}{dr}\right)\left[\frac{1}{r^2}\frac{d}{dr}\left(r^2\frac{d\Phi}{dr}\right) - 4\pi G_0\rho_0\right] = S \, ,
\label{np1}
\end{equation}
where $S=4\pi G_0\rho_0 a_0 = {\rm constant}$ inside the sphere. For the potential, we substitute
\begin{equation}
\Phi = \Phi_N(r) + \varphi(r) \, ,
\label{np2}
\end{equation}
where $\Phi_N$ is the Newtonian potential generally satisfying
\begin{equation}
\nabla^2\Phi_N = 4\pi G_0\rho_0\, ,
\label{newt}
\end{equation}
and $|\boldsymbol{\nabla}\Phi_N| \gg |\boldsymbol{\nabla}\varphi|$. To first order in $\varphi(r)$, eq.~(\ref{np1}) reduces to
\begin{equation}
\frac{d}{dr}\left(r^2\frac{d\varphi}{dr}\right) = 3\,a_0\,r \, .
\label{np3}
\end{equation}
The only particular solution of this Cauchy-Euler equation that remains finite at the origin and obeys the boundary condition $\varphi(0)=0$ is
\begin{equation}
\varphi(r) = \frac{3}{2}a_0\,r \, .
\label{np4}
\end{equation}
The Newtonian potential is determined by Gauss's law and the boundary condition $\Phi_N(0)=0$, viz.
\begin{equation}
\frac{d\Phi_N}{dr} = \frac{4\pi G_0\rho_0}{3}r ~~{\rm and}~~ \Phi_N(r)=\frac{2\pi G_0\rho_0}{3}r^2\, ,
\label{np5}
\end{equation}
so that the first-order solution~(\ref{np2}) can be written as
\begin{equation}
\Phi(r)=\frac{2\pi G_0\rho_0}{3}r^2 + \frac{3}{2}a_0\,r \, .
\label{np6}
\end{equation}
The first-order correction $(3a_0r)/2$ to the dominant Newtonian term is a linear term such as the $\gamma\,r$ term appearing in Weyl gravity. In our case, this term comes from $G(a)$ and appears both in the exterior solution (eq.~(\ref{weyl2})) and the interior solution (eq.~(\ref{np6})). \cite{man94} related that the linear term of Weyl gravity behaves in the same manner.

\label{lastpage}

\end{document}